# Observation of ballistic-diffusive thermal transport in GaN transistors using thermoreflectance thermal imaging

**Zhi-Ke Liu, Yang Shen, Han-Ling Li and Bing-Yang Cao***

Z. -K. Liu, Y. Shen, H. -L. Li, B. -Y. Cao*

Key Laboratory for Thermal Science and Power Engineering of Ministry of Education, Department of Engineering Mechanics, Tsinghua University, Beijing 100084, China
email: caoby@tsinghua.edu.cn

**Abstract** To develop effective thermal management strategies for gallium-nitride (GaN) transistors, it is essential to accurately predict the device junction temperature. Since the width of the heat generation in the devices is comparable to phonon mean free paths (MFPs) of GaN, phonon ballistic transport exists and can significantly affect the heat transport process, which necessitates a thorough understanding of the influence of the phonon ballistic effects in GaN transistors. In this paper, the ballistic-diffusive phonon transport in GaN-on-SiC devices is examined by measuring the hotspot temperature using the thermoreflectance thermal imaging (TTI) combined with the hybrid phonon Monte Carlo-diffusion simulations. A series of Au heaters are fabricated on the top of the GaN layer to quantitatively mimic the different heat source distributions during device operation. The experimental and simulation results show a good consistency and both indicate that the phonon ballistic effects can significantly increase the hotspot temperature. With the size of the heat source decreasing, the errors of Fourier's law-based predictions increase, which emphasizes the necessity to carefully consider the phonon ballistic transport in device thermal simulations.

## 1. Introduction

Gallium-nitride (GaN) high electron mobility transistors (HEMTs) are attractive devices for high-voltage and high-frequency applications due to its high breakdown voltage and high electron mobility [1, 2]. However, owing to the high power density (> 40 W/mm) [3], GaN HEMTs hold very high junction temperatures [4] which leads to a severe thermal concern [5]. It has been demonstrated that a 10 °C increase in the junction temperature can make the reliability of electronic devices decrease by 50% [6, 7]. An accurate assessment of the junction temperature of GaN HEMTs is critical to predict the device lifetime and develop effective thermal management strategies [8].

In GaN HEMTs, most heat is generated on the top of the GaN layer and concentrated at the drain-side gate edge [9, 10] with a width of a few hundred nanometers. The characteristic size is comparable to mean free paths (MFPs) of phonons, which are the dominant heat carriers in GaN [11, 12]. Due to the lack of phonon scattering in the heat source region, Fourier's law of heat conduction becomes inapplicable and significantly underestimates the hotspot temperature [13-15]. Also, being the result of Joule heating, the heat generation in GaN HEMTs is highly bias-dependent, i.e., the heat source distributions can be quite different under different working conditions. To accurately predict the junction temperature, it is essential to clearly understand the heat source size dependence of phonon ballistic effects, and their influence on the temperature.

Experimental efforts have been made to study the heat source size-induced phonon ballistic transport, including the studies using pulsed/modulated laser thermoreflectance [16], ultrafast coherent soft X-rays diffraction [14], transient thermal grating [17], extreme ultraviolet thermal and acoustic nanometrology [18], etc. However, in these techniques only the strength of the phonon ballistic effect can be inferred indirectly, the hotspot temperature and the device temperature field cannot be measured, which partly restrict the applications of these results in real devices [19]. As a non-contact optical technique, thermoreflectance thermal imaging (TTI) can directly detect the surface temperature with a submicron spatial resolution [20-25]. Compared with the above-mentioned methods, the influence of the phonon ballistic transport on the hotspot temperature can be directly detected by TTI, which can provide a more intuitive understanding of the phonon ballistic transport in real devices [26-28].

In this work, the influence of the phonon ballistic transport on the hotspot temperature in GaN HEMTs is studied using TTI. A series of Au heaters with different widths are fabricated on the top of the GaN layer to quantitively mimic the different heat source distributions in GaN HEMTs. The temperature fields and the hotspot temperatures of different samples are directly measured by TTI. Hybrid Monte Carlo (MC)-diffusion simulations are also conducted to validate the experimental results. Both the experimental and simulation results indicate that with the decrease in the heat source size, the hotspot temperature are much higher than the Fourier's law-based predictions. It is emphasized that the phonon ballistic effects should be carefully considered to accurately predict the device junction temperature.

## 2. Experimental

### 2.1 Thermoreflectance thermal imaging

TTI is an experimental technique to measure the surface temperature field with a high spatial resolution up to hundreds of nanometers. The spatial resolution can reach 353 nm by a 530 nm light source and a 100 × objective lens with a numerical aperture of 0.75, which makes TTI a highly effective technique to detect the hotspot temperature in electronic devices. TTI deduces the temperature field by probing the variation of the surface reflectivity with respect to temperatures. Under common working conditions of electronic devices, the dependence of the relative change in reflectivity on temperatures is linear, which can be expressed as

$$\frac{\Delta R}{R} = \left(\frac{1}{R}\frac{\partial R}{\partial T}\right)\Delta T = C_{\text{TR}}\Delta T, \tag{1}$$

where $R$ and $\Delta T$ represent the reflectivity and the temperature rise of the material surface, respectively. $C_{\text{TR}}$ is the coefficient of thermoreflectance [29], which is related to the type of material, the wavelength of light, and the surface roughness of the sample [30]. The range of $C_{\text{TR}}$ is usually from $10^{-5}\,\text{K}^{-1}$ to $10^{-2}\,\text{K}^{-1}$ for common metals and semiconductors. In prior to measuring the device temperature, the values of $C_{\text{TR}}$ of different samples are calibrated with adequate data averaging to improve the signal to noise ratio of the experiments [26].

TTI experiment system includes five subsystems, as shown in Fig. 1. During the experiment, the excitation subsystem produces an excitation signal, which is collected by the device under test (DUT), then a constant light-emitting diode (LED) light is shined on the DUT. The incident light is irradiated to the DUT surface through the microscope subsystem, and the reflected light from the DUT is captured by a charge coupled device (CCD) camera with synchronous lock-in detection. The excitation signals can lead to the temperature rise of the DUT surface, then the changed light intensity with respect to the temperature changes can be detected by the CCD. The surface temperature distributions can be extracted from the variation of the light intensity and $C_{\text{TR}}$ of the corresponding material. The timing setting of the excitation signal, the synchronous control of incident light, and the data analysis are completed by automatic system.

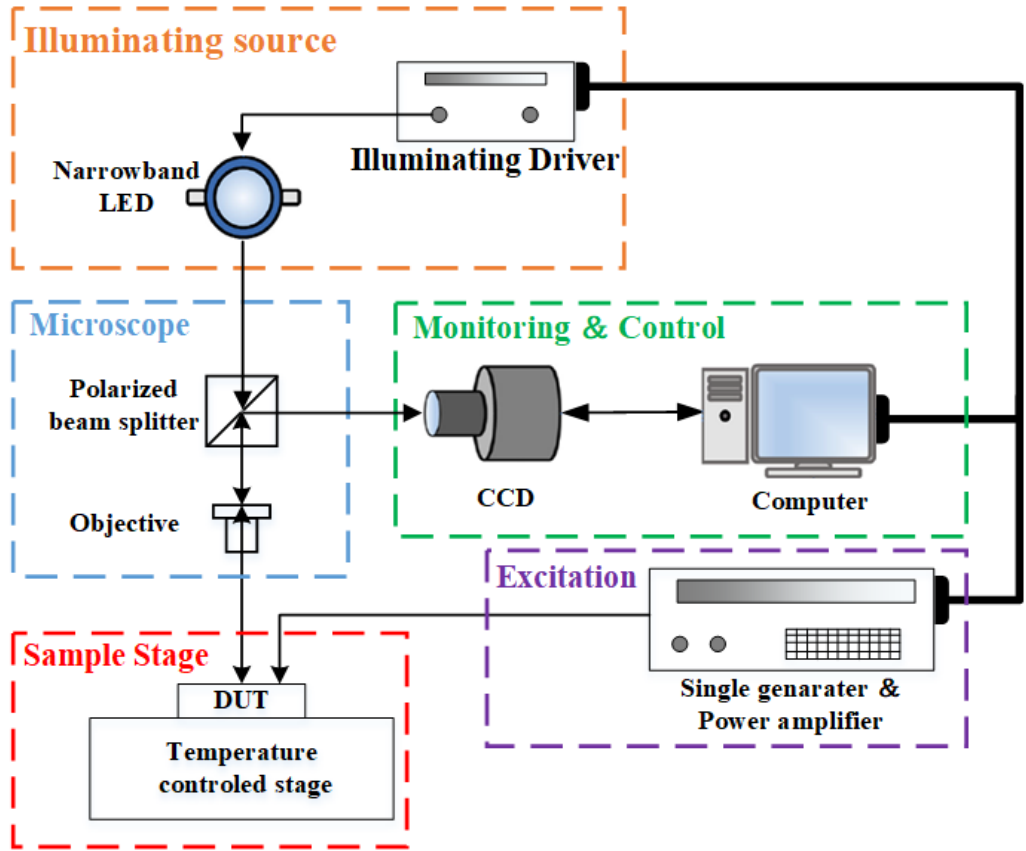

**Fig. 1** Schematic diagram of thermoreflectance thermal imaging

## 2.2 Device fabrication

The structure of the samples being measured is shown in Fig. 2, which consists of four layers. From top to bottom, they are a 25-nm $SiO_2$ passivation layer, 2.1-μm GaN buffer layer, a 50-nm AlN nucleation layer, and a 350-μm SiC substrate. The width and the thickness of the sample are $W = 7.5$ mm and $L = 7.5$ mm, respectively. Au heaters were fabricated on the top of the samples through electron beam lithography, metallization, and lift-off process. The width of the Au heaters ($w_g$) in different samples are 500 nm, 1 μm, 2 μm, 5 μm, and 10 μm. For the convenience of electrical excitation and electrical measurement of the heater resistance with the four-wire method, four contact pads in the size of $300 \times 300$ μm² were fabricated for each heater. Figure 3 shows the sample surface taken by the CCD. The white area is a pair of heaters with four pads, and the gray part is the top surface of the GaN layer.

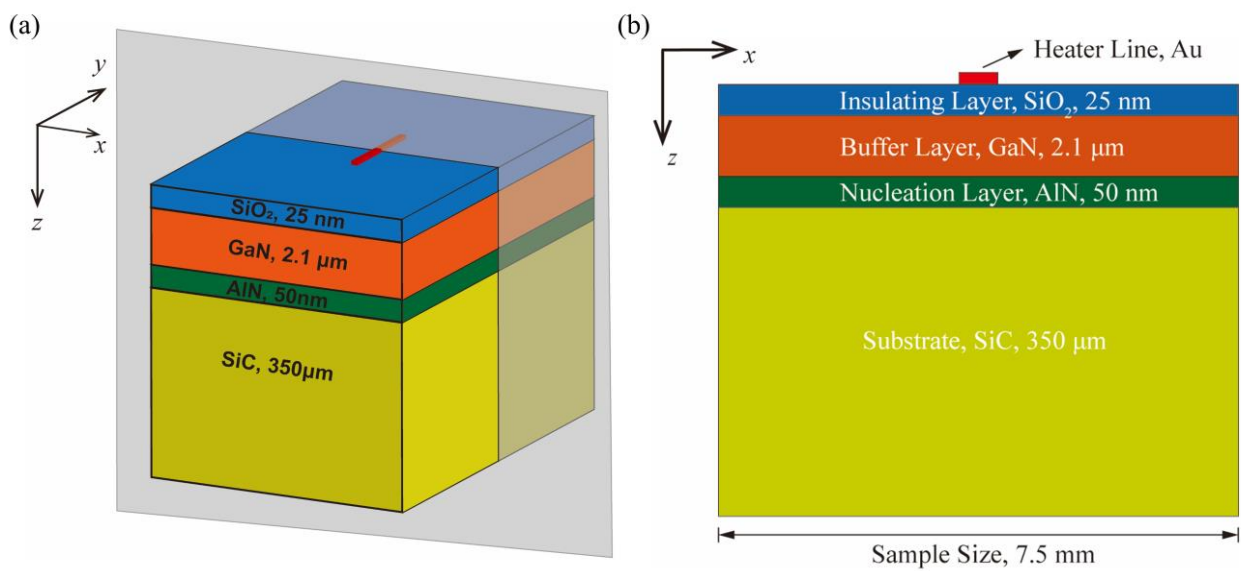

**Fig. 2** Sample structure diagram of (a) three-dimensional structure and (b) cross section

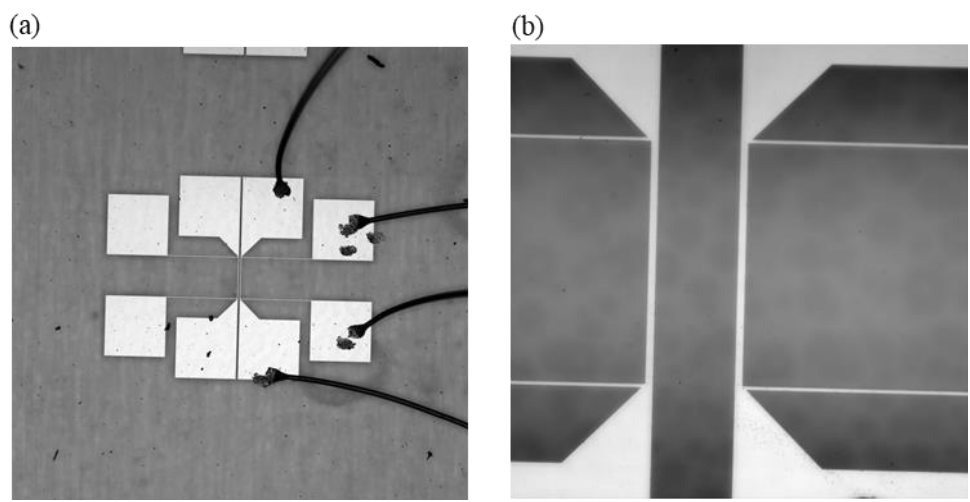

**Fig. 3** Micrograph of the upper surface of the sample taken by the CCD by (a) 5 × objective lens and (b) 60 × objective lens

## 3. Results and discussion

### 3.1 Measurement results of TTI

To measure the heat source temperatures of the samples, the 530 nm visible green LED light source is selected since it is sensitive to the temperature changes of Au [31, 32]. The coefficient of thermoreflectance $C_{TR}$ was determined by detecting the variation of the reflectivity with respect to a fixed 20 °C temperature difference applied to the samples, as indicated in Eq. (1). The measured $C_{TR}$ of Au of different samples are all $-2.8 \times 10^{-4}$ K$^{-1}$. Figure 4 shows the $C_{TR}$ distributions of the sample with $w_g = 2$ μm. It can be found that $C_{TR}$ in the Au region is uniform and constant, which indicates the effectiveness of the measurements.

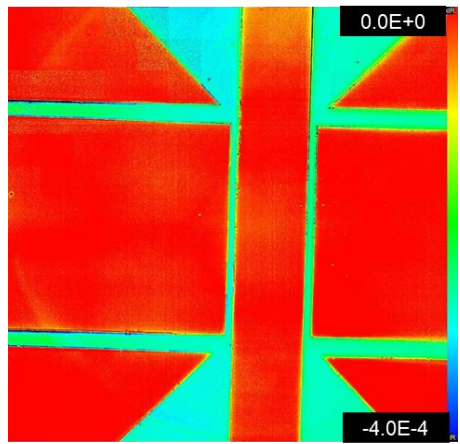

**Fig. 4** Distributions of $C_{TR}$ on the surface of the sample with $w_g = 2$ μm, in which the green area represents the Au material

With the $C_{TR}$ determined, the hotspot temperature rises of the samples with different power dissipations can be measured. The heating power is controlled by the input signal of the electrical excitation. Figure 5 shows the two-dimensional temperature rise distributions of the samples with different $w_g$. It can be found that the temperature rise mainly exists in the Au region and decays rapidly away from the heat source. Therefore, the average temperature rise in the Au region is defined as the hotspot temperature rise in this work. Figure 6 shows the hotspot temperature rise varying with the power dissipation of the sample with $w_g = 5$ μm. A good linear dependence is achieved the fitted thermal resistance $R_{TTI}$ is 75.52 K/W. The experiments on other samples also show a similar linearity.

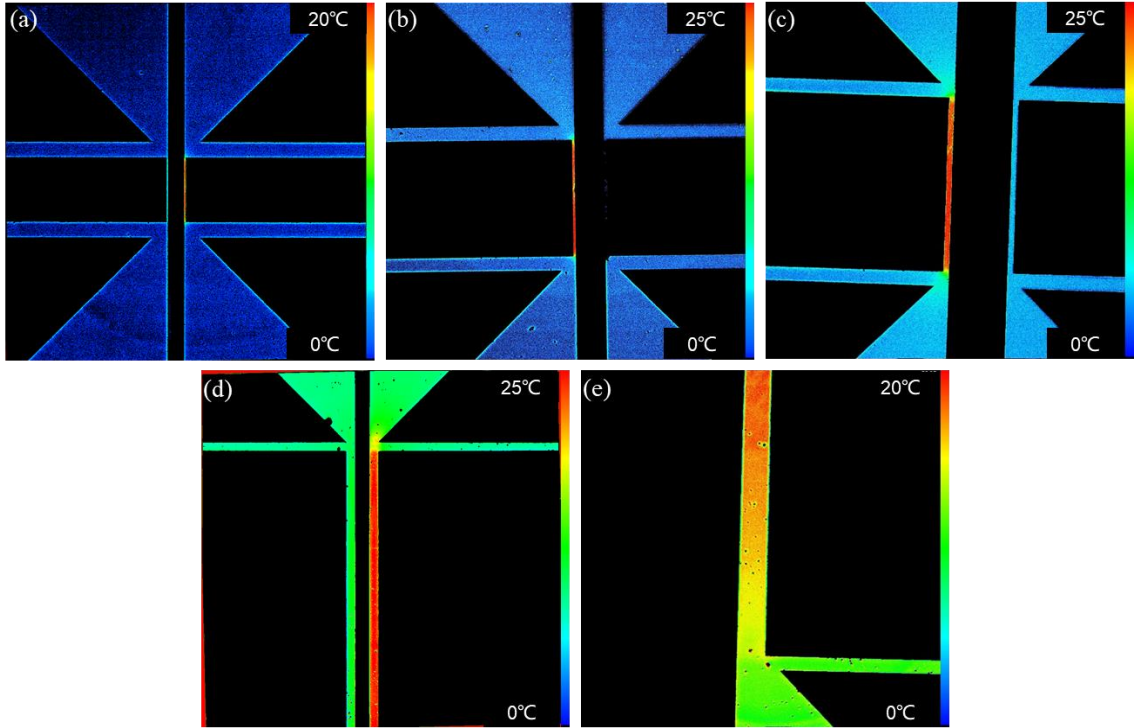

**Fig. 5** Two-dimensional maps of temperature rises in GaN samples heated by Au heaters with different $w_g$ (a) 500 nm, (b) 1 μm, (c) 2 μm, (d) 5 μm, (e) 10 μm

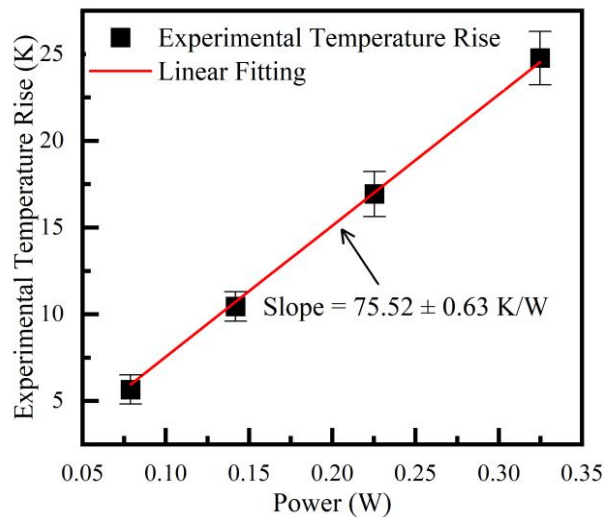

**Fig. 6** Linear dependence of the hotspot temperature rise measured by TTI on the power dissipation. The slope of the curve is the device thermal resistance

Figure 7 shows the measured thermal resistance $R_{TTI}$ varying with $w_g$. The Fourier's law-based FEM-calculated thermal resistance $R_{FEM}$ and the ratio between $R_{TTI}$ and $R_{FEM}$ are also given for comparison. The thermal conductivities of GaN and SiC are set as 168 W/(m·K) [33] and 350 W/(m·K) [34], respectively. As shown in Fig. 7, both $R_\mathrm{TTI}$ and $R_\mathrm{FEM}$ increases rapidly with $w_g$ decreasing, which can be attributed to the enhancement of the thermal spreading effect [35]. When heat spreads from a small source to a much larger region, there is a significant thermal spreading resistance, which increases with the decreased heat source size [36]. However, it can be noted that when $w_g$ is small enough, i.e., $w_g < 1$ μm, $R_\mathrm{TTI}$ becomes larger than the FEM-based predictions, and the deviation is enlarged with the decrease in $w_g$. This indicates that with the decrease in the heat source size, the phonon ballistic effect is enhanced which leads to a much higher junction temperature.

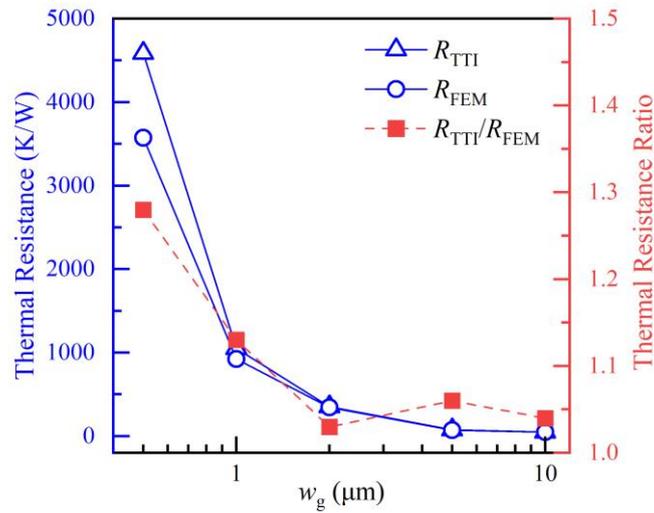

**Fig. 7** The TTI and FEM results of thermal resistance varying with the size of the heaters. The thermal resistance ratio is defined as the ratio between the thermal resistance predicted by TTI and FEM

### 3.2 Comparison with hybrid phonon MC-diffusion method

To illustrate the influence of the phonon ballistic effects on the thermal resistance more clearly, hybrid phonon MC-diffusion simulations are carried out for comparison with the experimental results [37, 38]. Considering the phonon ballistic effects mainly exist around the boundaries and the heat source, in the hybrid simulation the total device is divided into three parts: (1) the top MC zone that covers the entire GaN layer and a region extending from the GaN/SiC interface; (2) the bottom MC zone that covers the bottom boundary; and (3) the middle diffusion zone. Phonon MC simulations [39] are conducted only in the MC zone, and FEM simulations are carried out in the diffusion zone. There is an overlap region between the MC zone and adjacent diffusion zone, which is used for the information transfer and convergence check.

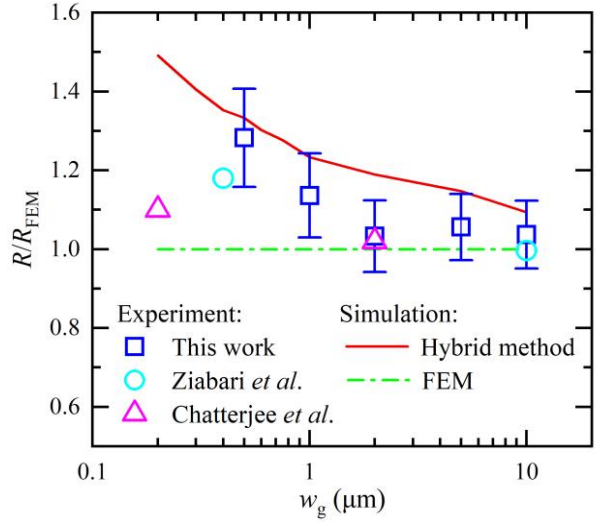

**Fig. 8** Experimental and simulated thermal resistance ratios varying with $w_g$. Reference values are extracted from Refs. [19, 40]

The results of FEM based on Fourier's law $R_{\text{FEM}}$ are used as reference values to calculate the thermal resistance ratio $R/R_{\text{FEM}}$ obtained by TTI measurements and hybrid phonon MC-diffusion simulations respectively, and Fig. 8 shows $R/R_{\text{FEM}}$ varying with the heat source size. Some values extracted from the literatures are also given for comparison. The influence of the thermal spreading effect is eliminated in $R/R_{\text{FEM}}$, thus, the ratio can reflect the impact of the phonon ballistic effects on the device thermal resistance. In Fig. 8, it can be found that $R/R_{\text{FEM}}$ increases with the decreased $w_g$, which implies the enhancement of the phonon ballistic effects. In the near-junction heat transport process [37], two kinds of phonon ballistic effects exist: one is the cross-plane ballistic effect caused by the phonon-boundary scattering, which is determined by the thickness of the GaN thin film. The other is the ballistic effect with the heat source size comparable to the phonon MFP. Since the thickness-dependent thermal conductivity of the GaN film is adopted in the FEM model, the deviation between $R/R_{\text{FEM}}$ and 1 can reflect the influence of the heat source size-induced ballistic effect on the thermal transport process. The results indicate that the ballistic effect with the heat source size comparable with the MFP is significantly enhanced with the decreased heat source size. It is illustrated that only using thickness-dependent thermal conductivities in FEM simulations will underestimate the phonon ballistic effect, which can lead to a lower junction temperature. The underestimation can increase with the decrease of the heat source size, e.g., at a negative bias when the heat is highly concentrated at the drain-side gate edge in GaN HEMTs [40]. It is emphasized that a multiscale thermal simulation is essential to accurately predict the device junction temperature.

## 4. Conclusions

In this work, the temperature field of the GaN devices are measured by TTI. Au heaters with different widths are fabricated on the top of the GaN layer to mimic the different heat source distributions during device operation. The thermal resistance of the different samples is deduced by conducting experiments with a series of power dissipations. It is found that with the decrease in the heat source size, the measured thermal resistance becomes higher than Fourier's law-based predictions, which can be attributed to the phonon ballistic effects. Hybrid phonon MC-diffusion simulations are carried out to investigate the influence of the phonon ballistic transport. It is illustrated that with the decrease in the size of the heat source, the ballistic effect with the heat source size comparable with phonon MFP is significantly enhanced which leads to a much higher hotspot temperature. The results emphasize the necessity of multiscale simulations and the need of high-resolution experiments to accurately predict the device junction temperature.

## Acknowledgements

This work was financially supported by National Natural Science Foundation of China (Grant Nos. U20A20301, 51825601)

## Declarations

**Competing Interests:** The authors declare that they have no conflict of interest.